A white paper for the ESO Expanding Horizons initiative

# Designing the Community Infrastructure for ESO's Next Transformational Facility
*Equitable Governance and Sustainable Teams for 2040s Astronomy*

Annagrazia Puglisi[1]*, Amelia Bayo[2], Laurane Fréour[3], Daniela Iglesias[4], Akhil Krishna[5], Fatemeh Zahra Majidi[6], Umberto Rescigno[7], Sabine Thater[3], Laurence Tresse[8], Tanya Urrutia[9]

The scientific ambitions of the 2040s will require large, interdisciplinary teams operating across continents, institutions, and increasingly heterogeneous political and funding landscapes. While significant effort is devoted to advancing the technical capabilities of future astronomical facilities, frameworks for coordinating and sustaining the associated community systems are often developed in parallel rather than embedded as coherent, long-term structures at the scale needed to fully realise this ambition. In this white paper, submitted as part of the ESO Expanding Horizons initiative, we draw on experience from established observatories and emerging collaborations to identify key community-level challenges. We argue that a central and transversal scientific challenge for the 2040s is to operate a flagship observatory in which access to telescope time, data, leadership, training, and career development is equitable across institutions, member states, and beyond. We propose that access and participation be treated as integral design parameters, embedded from the conceptual stage and sustained throughout the facility lifecycle, in order to ensure long-term scientific excellence, sustainability, and societal return.

***Author contributions.*** *Following the lead author, co-authors are listed alphabetically by family name. All co-authors contributed equally to the discussion and revision of the manuscript.*

Community feedback is welcomed via email (details below), and those who wish may indicate voluntary endorsement of the principles discussed via a short online form** linked below (see also page four of this document)

[1] School of Physics and Astronomy, University of Southampton, Highfield, SO17 1BJ, UK
[2] European Southern Observatory, Garching bei München, Karl-Schwarzschild-Straße 2, 85748 Garching bei München, Germany
[3] University of Vienna, Türkenschanzstraße 17, 1180 Vienna, Austria
[4] University of Leeds, Leeds LS2 9JT, UK
[5] Indian Institute of Astrophysics, 2nd Block, Koramangala, Bengaluru 560034, India
[6] INAF Osservatorio Astronomico di Capodimonte, Salita Moiariello 16, 8013, Napoli, Italy
[7] University of Atacama, Av. Copayapu 485, Copiapó, Atacama Region, Chile
[8] Aix Marseille Univ, CNRS, CNES, LAM, 38 rue Frédéric Joliot-Curie, 13388 Marseille Cedex 13, France
[9] Leibniz Institute for Astrophysics Potsdam (AIP), An der Sternwarte 16, 14482 Potsdam, Germany

*E-mail: a.puglisi@soton.ac.uk
**Link to endorsement form:
https://docs.google.com/forms/d/e/1FAIpQLSc6B-7kSdwb7v4eyi5zRAVbgUZ3Y43dpPrTCNfPHff6Hm4PWg/viewform?usp=header

In earlier observational eras, transformative breakthroughs could be achieved by small teams working at local observatories. By the late 20th century, the ambition of modern astronomical facilities demanded large collaborations involving dozens of institutions and countries to sustain costs, access to high-quality astronomical sites, and to provide the skilled workforce required to build these infrastructures and maximise their scientific return. Science teams have expanded accordingly: major observational programmes now operate through collaborations of hundreds across dozens of institutes worldwide, with some recent discoveries involving coordinated efforts of several thousand authors[1].

This trajectory makes clear that the scientific ambitions of the 2040s will demand interdisciplinary teams coordinating across continents, cultural and institutional environments, and within increasingly heterogeneous political and funding landscapes. Yet, the current context already reveals structural imbalances: gender representation remains uneven across career stages[2], cultural and geographic diversity is limited[3], and the communities closest to major observatories are often underrepresented in leadership, decision-making, and exploitation even where targeted programmes exist (e.g., ESO's Comite-Mixto funds for Antofagasta and Coquimbo). To deliver on the ambitions of the 2040s, community frameworks that allow large teams to thrive need to be embedded at the conceptual stage of a facility. This ensures that the fundamental scientific questions of the coming decades can be addressed by the full breadth of available talent, and that ESO's long-term investment delivers lasting scientific and societal return.

**Challenges and lessons from existing and emerging collaborations**

Over the past decade, projects such as SDSS, as well as facilities like ALMA, have shown that transformative science is achievable within a multi-institutional context, while broadening access to a wider scientific community. At the same time, the development history of facilities such as *JWST* highlights how exceptional scientific promise can be at risk when governance, coordination, or cost control are limited. At the data and infrastructure level, frameworks such as the FAIR principles demonstrate how shared standards can enable international collaboration. What remains largely missing are equivalent frameworks for coordinating and supporting the community systems needed to match technical ambition, enable belonging and participation, and ensure long-term sustainability[4]. Lessons from established observatories and emerging collaborations reveal several challenges:

1. **Structural and personal barriers to meaningful participation:** Junior scientists, caregivers (including but not limited to those with parental duties), and colleagues facing health, disability, mobility, socio-economic, or identity-based barriers (e.g. related to ethnicity, race, gender, or LGBTQIA+ identity) encounter obstacles such as incompatible meeting times, opaque structures, and fragmented communication channels, as well as unsafe legal or political environments. These factors shape who can relocate, attend meetings, or work at remote sites, and can lead to disconnection and uneven participation.

2. **Uneven institutional capacity:** Researchers at smaller institutions often lack the critical mass, technical support, or resources required to join leading collaborations thus concentrating expertise in a few already well-resourced centres.

3. **Precarity and career-stage vulnerability**: Scientists on short-term contracts must choose between investing in future facilities and protecting their immediate employability. Either choice represents a human cost, as well as a direct loss of scientific potential, expertise, and continuity.

4. **Limited engagement with local and Indigenous communities:** Astronomical facilities are often located in regions of deep cultural, environmental, and economic significance for local and Indigenous communities. When these communities are not meaningfully included in decision-making, such projects can unintentionally cause harm, limiting local benefits and undermining trust and long-term project viability, as illustrated by the long-standing debates around telescopes' construction in Hawaiʻi.

5. **Undervalued EDI and community work:** Much of the work required to sustain a functional scientific community (e.g., mentoring, access improvements, conflicts navigation, codes of conduct, ombudspersons' activity, and safe-space creation) is performed on top of scientific duties. Furthermore, these responsibilities are often taken by early-career researchers and/or individuals with lived experience of marginalisation. This is time-consuming, emotionally demanding, and rarely recognised in hiring, evaluation, or time-allocation systems. Without institutional support, redistribution and scientific reward, this creates a self-reinforcing cycle that hinders scientific progress, limits access to leadership positions[5], and drives away valuable community contributors.

Taken together, these experiences show that community capacity, encompassing psychological safety, equitable participation, sustainable workloads, trust and transparency in governance, and a shared sense of belonging, is a fundamental limiting factor in scientific output. Yet, they also indicate that sustainable systems are possible. Inclusive environments where early-career researchers and colleagues from marginalised groups are trusted with leadership and supported by positive role models consistently produce more ambitious, creative, and impactful science[6]. Concrete examples already exist: the LSST/Rubin Discovery Alliance illustrates how early integration of inclusive governance, open communication channels, and shared leadership broadens participation and accelerates scientific output. Similarly, *Roman* operates as an intentionally transparent, globally open network, with voluntary participation, open membership calls, and leadership positions filled through competitive, community-driven processes. Together, these models show that programmes prioritising mentoring networks, clear role assignment, and early-career leadership achieve higher engagement, faster mobilisation, and more diverse scientific contributions.

**From Challenges to Design Principles for ESO 2040**

We propose that a key scientific challenge for the 2040s is to operate a flagship facility in which access to telescope time, data, leadership, and training is equitable across institutions, member states and beyond, enabling a diverse global workforce to fully exploit its capabilities over decades. *This challenge is unique in its transversal nature: it affects every scientific theme, every working group, and every phase of a facility's lifecycle.* Addressing it requires considering access and participation as integral elements of the facility's design. To support this, we suggest that the next facility's design incorporates:

1. **Equitable access, redistribution and rewards mechanisms:** Explicitly accounting for differences in institutional capacity in time-allocation and data-access policies can create meaningful entry points for smaller or less-resourced institutes. Effective models include co-PI structures, mentoring partnerships, training and onboarding programmes that lower barriers for researchers and technical staff engaging with complex collaborations, instruments, pipelines, and surveys. Large astronomical projects also depend on extensive technical and analytical labour which can easily become invisible without explicit structures for attribution such as the historical case of the Harvard Computers[7]. Ensuring that software development, pipeline work, and other technical expertise are properly credited is therefore crucial for equitable access and participation, retention, and long-term scientific continuity.

2. **Transparent governance and equitable access to leadership roles:** Clear expectations around open calls, transparent criteria, wide advertisement, fixed terms, and predictable rotation of leadership roles can help prevent leadership from concentrating within a small set of institutions or demographic groups. Shadowing, mentoring, and co-leadership models can provide structured pathways for early-career researchers, scientists from smaller institutes, and colleagues from under-represented groups to enter leadership roles and gain visibility. Embedding these principles into consortium governance strengthens accountability, broadens participation, and improves scientific decision-making.

3. **Integrated EDI and community-governance structures:** Clearly defined and resourced EDI structures, with explicit mandates, reporting lines, and mechanisms for safe reporting (e.g. ombudspersons), are essential to support consortia working with the facility. Predictable resources, including budget lines for training, community activities, and professional support staff, together with EDI leadership that is diverse across geography, institution type, career stage, and multiple intersecting identities, are key to their effectiveness. Formal recognition of these contributions within service-time frameworks, proposal evaluations, promotions, and internal reporting can help ensure that community responsibilities are valued, particularly when accompanied by protected fractions of paid time. Rotating and distributing these roles can reduce the burden on a small number of individuals while building community leadership capacity across the collaboration. In addition, making the scientific achievements of those contributing to EDI and community work visible can help preventing the "EDI penalty", while sustained support can underpin ongoing skills development and access to high-quality training across the collaboration.

4. **Accountability, transparency, and community-health data:** Regular, centrally organised surveys on climate, inclusion, and access, implemented with strong ethical and data-protection safeguards, can provide valuable insight into community wellbeing and help identify emerging inequities. Transparent reporting on leadership demographics and participation patterns can further support accountability, particularly when paired with mechanisms that enable timely responses to identified disparities.

5. **Engagement with local and Indigenous communities:** Building dialogue with Indigenous and local communities into the facility's planning and development from the outset is an ethical responsibility and a foundation for long-term resilience. It is also a core component of maintaining the facility's social licence to operate, ensuring that environmental, cultural, and economic impacts are acknowledged and addressed collaboratively. Involving local communities and relevant stakeholders in decision-making, for example around choices of energy sources for telescope operations, can foster co-benefits and lead to more socially and environmentally sustainable solutions[8].

6. **Open and transparent communication with the wider community:** Clear and accessible communication channels between ESO, consortia, and the broader community are central to ensuring that diverse perspectives can be heard and included. As no single mechanism can fully reflect the range of experiences across identities, states and institutions, offering multiple and complementary modes of engagement (e.g., open consultations, dedicated forums for marginalised or underrepresented groups), can facilitate the meaningful inclusion of community voices in both governance and facility evolution.

These principles are strongly aligned with evolving expectations across the European research landscape and fully support ESO's Expanding Horizons vision for socially sustainable infrastructure in the 2040s. By addressing these structural challenges at the onset, the next-generation facility can broaden participation, retain talent, and enable a more resilient and creative global community, ensuring that ESO's investment delivers its full scientific and societal potential. As a flagship organisation operating across borders and grounded in European values, ESO is uniquely positioned to pioneer these standards. Recent experiences from initiatives such as Rubin/LSST illustrate both the promise and the fragility of EDI efforts in today's volatile national political climates, underscoring the need for such commitments to be embedded structurally rather than left to voluntary or local initiatives.

**Facility and technology implications**

The facility and technology implications lie in the infrastructures that enable large, equitable collaborations to function effectively. Any next-generation ESO programme would benefit from the following community-enabling elements to help fulfil its scientific promise in the 2040s:

- Integrated collaboration platforms that centralise communication, documentation, training materials, and decision-making processes, reducing fragmentation across working groups and ensuring that new members can onboard efficiently.
- Data systems that build on positive examples such as existing ESO and ALMA archives, supporting fair and effective access and providing targeted technical support for smaller or less-resourced institutes working with large and complex data sets.
- Survey and analytics tools that allow high-level monitoring of community health, participation patterns, and accessibility barriers in a privacy-respecting and GDPR-compliant way, providing evidence to inform governance decisions and identify emerging issues early.

**Community feedback and endorsement (arXiv version only).** This white paper has been submitted to the ESO Expanding Horizons initiative and is posted on arXiv as a starting point for constructive discussion on how community, governance, and equity considerations can be integrated into the design of future large astronomical facilities. We acknowledge that this perspective reflects our own experiences and may be incomplete, and we therefore welcome the inclusion of additional voices.

To encourage broader input, we invite members of the astronomy and astrophysics community to share feedback and comments with the lead authors. Input from a broad and diverse community will help ensure that the perspectives and challenges discussed here are not limited to any single viewpoint.

Those who wish to indicate their endorsement of the general principles outlined here may also do so via a short, voluntary form available at: [link]. Visible engagement from across the community can help amplify this conversation and ensure it reflects a broader range of experiences. Endorsements are intended to reflect interest in and support for the overall direction of the discussion, rather than agreement with every specific recommendation, and are not to be interpreted as institutional or formal endorsements.

The views expressed in the white paper reflect those of the authors and do not necessarily represent the views of their affiliated institutions or organisations.